\documentclass[12pt]{article}
\begin{document}
%\draft
\title
{A fine quantum mechanism of classical gravity}
\author
{By Michael A. Ivanov \\
Chair of Physics, \\
Belarus State University of Informatics and Radioelectronics, \\
6 P. Brovka Street,  BY 220027, Minsk, Republic of Belarus.\\
E-mail: ivanovma@gw.bsuir.unibel.by.}

\maketitle

\begin{abstract}
It is shown that screening the background of super-strong
interacting gravitons ensures the
Newtonian attraction, if a part of single gravitons is pairing and
graviton pairs are destructed by collisions with a body. If the
considered quantum mechanism of classical gravity is realized in
the nature, than an existence of black holes contradicts to
the equivalence principle. In such the model, Newton's
constant is proportional to $H^{2}/T^{4},$ where $H$ is the Hubble
constant, $T$ is an equivalent temperature of the graviton
background. The estimate of the Hubble constant is obtained for the
Newtonian limit:
$H=3.026 \cdot 10^{-18} \  s^{-1}$ (or $94.576 \  km \cdot s^{-1}
\cdot Mpc^{-1}$).
\end{abstract}

\section[1]{Introduction }
It was shown by the author in the previous study \cite{1,2} that
an alternative explanation of cosmological redshift as a result of
interaction of a photon with the graviton background is possible.
In the case, observed dimming of supernovae Ia \cite{3} and the
Pioneer 10 anomaly \cite{4} may be explained from one point of
view as additional manifestations of interaction with the graviton
background. Some primary features of a new cosmological model,
based on this approach, are described in author's preprint
\cite{4a}.
\par
In this study (for details, see author's full paper \cite{20}),
forces of gravitonic radiation pressure are
considered which act on bodies in a presence of such the
background. It is shown that pressure of single gravitons of the
background, which run against a body pair from infinity, results
in mutual attraction of bodies with a magnitude which should be
approximately $1000$ times greater than Newtonian attraction. But
pressure of gravitons scattered by bodies gives a repulsion force
of the same order; the last is almost exact compensating this
attraction. To get Newton's law of gravity, it is necessary to
assume that gravitons form correlated pairs. By collision with a
body, such a pair should destruct in single gravitons. Flying away
gravitons of a pair should happen in independent directions, that
decreases a full cross-section of interaction with scattered
gravitons. As a result, an attraction force will exceed a
corresponding repulsion force acting between bodies. In such the
model, Newton's constant is connected with the Hubble constant
that gives a possibility to obtain a theoretical estimate of the
last. We deal here with a flat non-expanding universe fulfilled
with superstrong interacting gravitons; it changes the meaning of
the Hubble constant which describes magnitude of three small
effects of quantum gravity but not any expansion.
\par
\section[2]{Screening the graviton background}

\par
If gravitons of the background run against a pair of bodies with
masses $m_{1}$ and $m_{2}$ (and energies $E_{1}$ and $E_{2}$) from
infinity, then a part of gravitons is screened. Let $\sigma
(E_{1},\epsilon)$ is a cross-section of interaction of body $1$
with a graviton with an energy $\epsilon=\hbar \omega,$ where
$\omega$ is a graviton frequency, $\sigma (E_{2},\epsilon)$ is the
same cross-section for body $2.$ In absence of body $2,$ a whole
modulus of a gravitonic pressure force acting on body $1$ would be
equal to:
\begin{equation}
4\sigma (E_{1},<\epsilon>)\cdot {1 \over 3} \cdot {4 f(\omega, T)
\over c},
\end{equation}
where $f(\omega, T)$ is a graviton spectrum with a temperature $T$
(assuming to be planckian), the factor $4$ in front of $\sigma
(E_{1},<\epsilon>)$ is introduced to allow all possible directions
of graviton running, $<\epsilon>$ is another average energy of
running gravitons with a frequency $\omega$ taking into account a
probability of that in a realization of flat wave a number of
gravitons may be equal to zero, and that not all of gravitons ride
at a body.
\par
Body $2,$ placed on a distance $r$ from body $1,$ will screen a
portion of running against body $1$ gravitons which is equal for
big distances between the bodies (i.e. by $\sigma
(E_{2},<\epsilon>) \ll 4 \pi r^{2}$):
\begin{equation}
\sigma (E_{2},<\epsilon>) \over 4 \pi r^{2}.
\end{equation}
Taking into account
all frequencies $\omega,$ an attractive force will act
between bodies $1$ and $2:$
\begin{equation}
F_{1}= \int_{0}^{\infty} {\sigma (E_{2},<\epsilon>) \over 4 \pi
r^{2}} \cdot 4 \sigma (E_{1},<\epsilon>)\cdot {1 \over 3} \cdot {4
f(\omega, T) \over c} d\omega.
\end{equation}
Let $f(\omega, T)$ is described with the Planck formula,
$x \equiv {\hbar \omega/  kT},$ and $\bar{n} \equiv {1/
(\exp(x)-1)}$ is an average number of gravitons in a flat wave
with a frequency $\omega$ (on one mode of two distinguishing with
a projection of particle spin). Let $P(n,x)$ is a probability of
that in a realization of flat wave a number of gravitons is equal
to $n,$ for example $P(0,x)=\exp(-\bar{n}).$
\par
If $P_{1}$ is a
probability that a single graviton will ride namely at the body
and one assume that
$P_{1}=P(1,x),$
where $P(1,x)=\bar{n}\exp(-\bar{n}),$ (below it is admitted for
pairing gravitons: $P_{1}=P(1,2x)$ - see section 4),
then the  quantity $<\epsilon>$ is equal to:
$<\epsilon>= \hbar \omega (1-P(0,x))\bar{n}^{2}\exp(-\bar{n}).$
Then we get for an attraction force $F_{1}:$
$F_{1}\equiv G_{1} \cdot  m_{1}m_{2}/r^{2}$ where the constant
$G_{1}$ is equal by $T=2.7 K$ to \cite{20}
$G_{1} =1215.4 \cdot G,$
that is three order greater than Newton's constant $G.$
\par
But if gravitons are elastic scattered with body $1,$ then our
reasoning may be reversed: the same portion (2) of scattered
gravitons will create a repulsive force $F_{1}^{'}$ acting on body
$2$ and equal to $F_{1}^{'} =F_{1},$
if one neglects with small allowances which are proportional to
$D^{3}/  r^{4}$  (see section 5).
\par
So, for bodies which elastic scatter gravitons, screening a flux
of single gravitons does not ensure Newtonian attraction. But for
black holes which absorb any particles and do not
re-emit them, we will have $F_{1}^{'} =0.$ It means that such the
object would attract other bodies with a force which is
proportional to $G_{1}$ but not to $G,$ i.e. Einstein's
equivalence principle would be violated for it. This conclusion,
as we shall see below, stays in force for the case of graviton
pairing too.
\section[3]{Graviton pairing}
To ensure an attractive force which is not equal to a repulsive
one, particle correlations should differ for {\it in} and {\it
out} flux. For example, single gravitons of running flux may
associate in pairs. If such pairs are destructed by collision with a
body, then quantities $<\epsilon>$ will distinguish for running
and scattered particles. Graviton pairing may be caused with
graviton's
own gravitational attraction or gravitonic spin-spin interaction.
\par
To find an average number of pairs $\bar{n}_{2}$ in a wave with a
frequency $\omega$ for the state of thermodynamic equilibrium, one
may replace $\hbar \rightarrow 2\hbar$ by deducing the Planck
formula. Then an average number of pairs will be equal to:
$\bar{n}_{2} ={1 / (\exp(2x)-1}),$
and an energy of one pair will be equal to $2\hbar \omega.$ It is
important that graviton pairing does not change a number of
stationary waves, so as pairs nucleate from existing gravitons.
The question arises: how many different modes, i.e. spin
projections, may have graviton pairs?  It follows
from the energy conservation law that composite gravitons should
be distributed only in two modes \cite{20}.
The spectrum of composite gravitons is proportional to the Planck
one;  an equivalent temperature of this sub-system is
$T_{2} \equiv (1/8)^{1/4}T =  0.5946T.$
It is important that the graviton pairing effect does not changes
computed values of the Hubble constant and of anomalous
deceleration of massive bodies \cite{1}: twice decreasing of a
sub-system particle number due to the pairing effect is
compensated with twice increasing the cross-section of interaction
of a photon or any body with such the composite gravitons.

\section[4]{Computation of Newton's constant}
If running graviton pairs ensure for two bodies an attractive
force $F_{2},$ then a repulsive force due to re-emission of
gravitons of a pair alone will be equal to $F_{2}^{'} =F_{2}/2.$
It follows from that the cross-section $\sigma (E_{2},<\epsilon>)=
{1 \over 2} \cdot \sigma (E_{2},<\epsilon_{2}>),$ where
$<\epsilon_{2}>$ is an average pair energy with taking into
account a probability of that in a realization of flat wave a
number of graviton pairs may be equal to zero, and that not all of
graviton pairs ride at a body ($<\epsilon_{2}>$ is an analog of
$<\epsilon>$). We get for graviton pairs:
\begin{equation}
<\epsilon_{2}> = 2\hbar \omega
(1-P(0,2x))\bar{n}_{2}^{2}\exp(-\bar{n}_{2}) \cdot P(0,x).
\end{equation}
Then a force of attraction of two bodies due to pressure of
graviton pairs $F_{2}$ will be equal to:
\begin{equation}
F_{2}= {4 \over 3} \cdot
{D^{2} c(kT)^{6} m_{1}m_{2} \over {\pi^{3}\hbar^{3}r^{2}}}\cdot
I_{2},
\end{equation}
where $I_{2} = 2.3184 \cdot 10^{-6}.$
The difference $F$ between attractive and repulsive forces will be
equal to:
\begin{equation}
F \equiv F_{2}- F_{2}^{'}={1 \over 2}F_{2} \equiv G_{2}{m_{1}m_{2}
\over r^{2}},
\end{equation}
where the constant $G_{2}$ is:
\begin{equation}
G_{2} \equiv {2 \over 3} \cdot {D^{2} c(kT)^{6} \over
{\pi^{3}\hbar^{3}}} \cdot I_{2}.
\end{equation}
As $G_{1}$ as well $G_{2}$ are proportional to $T^{6}$ (and $H
\sim T^{5}$).
\par
If one assumes that $G_{2}=G,$ then by
$T=2.7K$ the constant $D$ should have the value:
$D=1.124 \cdot 10^{-27}{m^{2} / eV^{2}}.$
We can establish a connection between the two
fundamental constants $G$ and $H$ under the condition that
$G_{2}=G:$
\begin{equation}
H= (G  {45 \over 32 \pi^{5}}  {\sigma T^{4} I_{4}^{2}
\over {c^{3}I_{2}}})^{1/2}= 3.026 \cdot 10^{-18}s^{-1},
\end{equation}
or in the units which are more familiar for many of us: $H=94.576 \
km \cdot s^{-1} \cdot Mpc^{-1}.$
\par
This value of $H$ is significantly larger than we see in the
majority of present astrophysical estimations \cite{3,12}, but it is
well consistent with some of them \cite{12a} and with the observed
value of anomalous acceleration of Pioneer 10 \cite{4} $w=(8.4 \pm
1.33)\cdot 10^{-10} \ m/s^{2}.$ Any massive body, moving relative
to the background, must feel a deceleration $w \simeq Hc$
\cite{1,2}; with $H=3.026 \cdot 10^{-18}s^{-1}$ we have $Hc=9.078
\cdot 10^{-10} \ m/s^{2}.$

\section[5]{Why and when gravity is geometry}
The described quantum mechanism of classical gravity gives
Newton's law with the constant $G_{2}$ value (7) and the
connection (8) for the constants $G_{2}$ and $H.$  We have
obtained the rational value of $H$  by $G_{2} = G,$ if the
condition of big distances is fulfilled:
\begin{equation}
\sigma (E_{2},<\epsilon>) \ll 4 \pi r^{2}.
\end{equation}
Because it is known from experience that for big bodies of the
solar system, Newton's law is a very good approximation, one would
expect that the condition (9) is fulfilled, for example, for the
pair Sun-Earth. But assuming $r=1 \ AU$ and
$E_{2}=m_{\odot}c^{2},$ we obtain assuming for rough estimation
$<\epsilon> \rightarrow
\bar{\epsilon}:$ ${\sigma (E_{2},<\epsilon>) / 4 \pi r^{2}}
\sim 4 \cdot 10^{12}. $ It means that in the case of interaction
of gravitons or graviton pairs with the Sun in the aggregate, the
considered quantum mechanism of classical gravity could not lead
to Newton's law as a good approximation. This "contradiction" with
experience is eliminated if one assumes that gravitons interact
with "small particles" of matter - for example, with atoms. If the
Sun contains of $N$ atoms, then $\sigma (E_{2},<\epsilon>)=N \sigma
(E_{a},<\epsilon>),$ where $E_{a}$ is an average energy of one
atom. For rough estimation we assume here that $E_{a}=E_{p},$ where
$E_{p}$ is a proton rest energy; then it is $N \sim 10^{57},$ i.e.
${\sigma (E_{a},<\epsilon>)/ 4 \pi r^{2}} \sim  10^{-45} \ll 1.$
\par
This necessity of "atomic structure" of matter for working the
described quantum mechanism is natural relative to usual
bodies. But would one expect that black holes have a similar
structure? If any radiation cannot be emitted with a black hole, a
black hole should interact with gravitons as an aggregated object,
i.e. the condition (9) for a black hole of sun mass has not been
fulfilled even at distances $\sim 10^{6} \ AU.$
\par
For bodies without an atomic structure, the allowances, which are
proportional to $D^{3}/ r^{4}$ and are caused by decreasing a
gravitonic flux due to the screening effect, will have a factor
$m_{1}^{2}m_{2}$ or $m_{1}m_{2}^{2}.$ These allowances break the
equivalence principle for such the bodies.
\par
For bodies with an atomic structure, a force of interaction is
added up from small forces of interaction of their "atoms": $$
F \sim N_{1}N_{2}m_{a}^{2}/r^{2}=m_{1}m_{2}/r^{2},$$ where
$N_{1}$ and $N_{2}$ are numbers of atoms for bodies $1$ and $2$.
The allowances to full forces due to the screening effect will be
proportional to the quantity: $N_{1}N_{2}m_{a}^{3}/r^{4},$ which
can be expressed via the full masses of bodies as
$m_{1}^{2}m_{2}/r^{4}N_{1}$ or $m_{1}m_{2}^{2}/r^{4}N_{2}.$ By big
numbers $N_{1}$ and $N_{2}$ the allowances will be small. The
allowance to the force $F,$ acting on body $2,$ will be equal to:
\begin{equation}
\Delta F = {1 \over 3N_{2}} \cdot {{D^{3} c^{3} (kT)^{7}
m_{1} m_{2}^{2}} \over {\pi^{4}\hbar^{3}r^{4}}} \cdot I_{3},
\end{equation}
(for body $1$ we shall have the similar expression if replace
$N_{2} \rightarrow N_{1}, \ m_{1}m_{2}^{2} \rightarrow
m_{1}^{2}m_{2}$), where $ I_{3} = 1.0988 \cdot 10^{-7}. $
\par
Let us find the ratio:
\begin{equation}
{\Delta F \over F} = {D E_{2} kT \over {N_{2} 2\pi r^{2}}} \cdot
{I_{3} \over I_{2}}.
\end{equation}
Using this formula, we can find by $E_{2}=E_{\odot}, \ r=1 \ AU:$
${\Delta F / F} \sim 10^{-46}.$
\par
An analogical allowance to the force $F_{1}$ has  the
order $\sim 10^{-48}F_{1},$ or $\sim 10^{-45}F.$
We see that for bodies with an atomic structure
the considered mechanism leads to very small deviations from
Einstein's equivalence principle, if the condition (9) is
fulfilled for microparticles, which prompt interact with
gravitons.
\par
For small distances we shall have:
\begin{equation}
\sigma (E_{2},<\epsilon>) \sim 4 \pi r^{2}.
\end{equation}
It takes place by $E_{a}=E_{p}, \ <\epsilon> \sim 10^{-3} \ eV$
for $r \sim 10^{-11} \ m.$ This quantity is many order larger than
the Planck length. The equivalence principle should be broken at
such distances.

\section[6]{Conclusion}
It is known that giant intellectual efforts to construct a quantum
theory of metric field, based on the theory of general relativity,
have not a hit until today (see \cite{18}). From
a point of view of the considered approach, one may explain it by
the fact that gravity is not geometry at short distances $\sim
10^{-11} \ m.$ Actually, it means that at such the distances quantum
gravity cannot be described alone.
\par
It follows from the present study that the geometrical
description of gravity should be a good idealization at big distances
by the
condition of "atomic structure" of matter. This condition cannot
be accepted only for black holes which must interact with
gravitons as aggregated objects. In addition, the equivalence
principle is roughly broken for black holes, if the described
quantum mechanism of classical gravity is realized in the nature.
\par
Other important features of this mechanism are the following ones.
$\\ \bullet$ Attracting bodies are not initial sources of
gravitons. In this sense, a future theory must be non-local to
describe gravitons running from infinity.
\\ $\bullet$ Newton's law takes place if gravitons are
pairing; to get preponderance of attraction under repulsion,
graviton pairs should be destructed by interaction with matter
particles. \\ $\bullet$ The described quantum mechanism of
classical gravity is obviously asymmetric relative to the time
inversion. By the time inversion, single gravitons would run
against bodies forming pairs. It would lead to replacing a body
attraction with a repulsion. But such the change will do
impossible graviton pairing.
\\ $\bullet$ The two fundamental constants - Newton's and Hubble's
ones - are connected with each other in such the model. The
estimate of Hubble's constant has been got here using an
additional postulate $P_{1}=P(1,2x)$ for pairing gravitons.
\\ $\bullet$ It is proven that graviton pairs should be
distributed in two modes with different spin projections.
\par
A future theory dealing with gravitons as usual particles having
an energy, a momentum etc
should have a number of features, which are not characterizing
any existing model, to image the recounted above features of a
possible quantum mechanism of gravity.

%\begin{references}

%\end{references}

\begin{thebibliography}{References                        }
\bibitem{1}
M.A.Ivanov, General Relativity and Gravitation, {\bf 33}, 479
(2001); Erratum: {\bf 35}, 939 (2003); [astro-ph/0005084 v2].
\bibitem{2}
M.A.Ivanov, [gr-qc/0009043]; Proc. of the Int. Symp. "Frontiers of
Fundamental Physics 4" (9-13 Dec 2000, Hyderabad, India), Eds B.G.
Sidharth and M.V. Altaisky, Kluwer Academic/Plenum Publishers,
August 2001; Proc. of the 4th Edoardo Amaldi Conference on
Gravitational Waves (Perth, Western Australia, 8-13 July 2001)
Class. Quantum Grav. {\bf 19}, 1351 (2002).
\bibitem{3}
A.G. Riess et al.   AJ {\bf 116}, 1009 (1998).
\bibitem{4}
Anderson, J.D. et al. Phys. Rev. Lett., 1998, v.81, p. 2858; Phys.
Rev. {\bf D65} (2002) 082004. [gr-qc/0104064 v4]
\bibitem{4a}
M.A.Ivanov. Model of graviton-dusty universe. [gr-qc/0107047]
\bibitem{20}
M.A.Ivanov. Screening the graviton background, graviton pairing, and
Newtonian gravity. [gr-qc/0207006]
\bibitem{12}
W. L. Freedman et al. ApJ, {\bf 553} (2001) 47.
\bibitem{12a}
J.A. Willick, Puneet Batra. ApJ, {\bf 548} (2001) 564.
\bibitem{18}
S. Carlip. Quantum gravity: a progress report. [gr-qc/0108040]
\end{thebibliography}
\end{document}